\documentclass[useAMS,usenatbib,usegraphicx]{mn2e}

\newcommand{\dd}{{\rm d}}
\newcommand{\Mp}{m_{\rm p}}

\title[SZ in the Virgo cluster]
{Sunyaev-Zel'dovich effect in the Virgo cluster from WMAP and ROSAT data.}

\author[J.~M.~Diego and Y.~Ascasibar]
{
  J.~M.~Diego$^{1,2}$\thanks{E-mail: jdiego@ifca.unican.es} and Y.~Ascasibar$^3$\\
  $^1$ IFCA, Instituto de F\'\i sica de Cantabria, Avda. Los Castros s/n. 39005 Santander, Spain\\
  $^2$ {\it Consejo Superior de Investigaciones Cient\'\i ficas}.\\
  $^3$ Astrophysikalisches Institut Potsdam, An der Sternwarte 16, 14482 Potsdam, Germany
}

\date{Draft version 9 (\today)}
\pagerange{\pageref{firstpage}--\pageref{lastpage}}

\begin{document}
\maketitle

\label{firstpage}


\begin{abstract}
WMAP observations at mm wavelengths are sensitive to the Sunyaev-Zel'dovich effect in galaxy clusters. Among all the objects in the sky, the Virgo cluster is expected to provide the largest integrated signal. Based on models compatible with the X-ray emission observed in the ROSAT All Sky Survey, we predict an approximately two-sigma detection of the SZ effect from Virgo in the WMAP 3-year data. Our analysis reveals a 3-sigma signal on scales of 5 degrees, although the frequency dependence deviates from the theoretical expectation for the SZ effect.
The main sources of uncertainty are instrumental noise, and most importantly, possible contamination from point sources and diffuse back/foregrounds.
In particular, a population of unresolved extragalactic sources in Virgo would explain the observed intensity and frequency dependence. 
In order to resolve this question, one needs to wait for experiments like Planck to achieve the required accuracy.

\end{abstract}



\section{Introduction}
\label{secIntro}

The X-ray emission and the Sunyaev-Zel'dovich (SZ) effect provide complementary probes of the physical state of the hot intracluster medium (ICM). In galaxy clusters, the dominant mechanism for X-ray emission is thermal bremsstrahlung, and the surface brightness is roughly proportional to the integral along the line of sight of the gas density squared \citep{Sarazin86}. The SZ effect, on the other hand, is proportional to the integrated energy density of the electrons, i.e. the gas pressure \citep{SunyaevZeldovich72,Birkinshaw99}. By combining X-ray and microwave observations it is possible to constrain the angular distance \citep[e.g.][]{Birkinshaw+91,Mason+01,Reese+02,Jones+05,Bonamente+06} and the baryon fraction \citep[e.g.][]{Myers+97,Grego+01,LaRoque+06} of galaxy clusters. Measuring these quantities for a substantial sample of objects provides a stringent benchmark to test cosmological models.

In general, there is good agreement between estimates of the gas density based on X-ray and interferometric radio data, at least within the central regions accessible to both types of observation.
Such agreement has also been reported at millimeter wavelengths, although the results seem to indicate a baryon fraction that is considerably lower than the cosmic value \citep{Afshordi+05,Afshordi+07,McCarthy+07}.
While this would be perfectly consistent with measurements based on X-ray observations only \citep[e.g.][]{EttoriFabian99,Roussel+00,Allen+02,Ettori03,Vikhlinin+06} as well as theoretical estimates of the baryon fraction expected in the central regions of the ICM \citep[e.g.][]{Ascasibar+03,Ascasibar+06_SR,Kravtsov+05,Ettori+06}, recent studies based on WMAP data have suggested that the SZ effect observed in galaxy clusters might actually be weaker than one would infer from their X-ray luminosity \citep{Lieu+06,BielbyShanks_07}.
The presence of point sources could compensate part of the SZ temperature decrement at WMAP frequencies, but the required flux would be much higher than the observational upper limits \citep{Coble+07,LinMohr07}.
An interesting alternative explanation would be that a non-thermal population of relativistic electrons promoted CMB photons into the X-ray regime by inverse Compton scattering \citep{LieuQuenby_06}.
In this scenario, the gas density required to produce the observed X-ray flux would be lower than for a thermalized plasma, and the corresponding SZ effect would be dimmer.

In this paper we address the question of whether the SZ effect detected in WMAP-3yr data is compatible with the observed X-ray emission. We focus on the brightest and most extended source in the sky at both wavelenghts: the Virgo cluster. Due to its small distance from us and its relatively large mass, the integrated SZ flux from Virgo must be larger than that from any other cluster \citep{Taylor+03}. This object has also the largest integrated X-ray flux, as can be clearly seen, for instance, in the data from the ROSAT All Sky Survey.
We expect the SZ effect from Virgo to extend further than the X-ray signal, covering more than a dozen square degrees. Although it should be a prominent source in WMAP data, a review of the literature shows a surprising lack of results on this matter, suggesting that previous attempts have not found a significant detection.

Here we investigate two different models of the intracluster medium that are consistent with the observed X-ray emission from Virgo, and we predict their expected SZ effect in WMAP data. Our analytical models and the theoretical prescription to estimate the X-ray surface brightness and the SZ effect are described in Section~\ref{secModel}.
Section~\ref{secXray} describes the analysis of ROSAT X-ray data.
We discuss the WMAP data in Section~\ref{secWMAP}, where we present our measurement of the SZ signal in Virgo and compare it with the estimates based on the X-rays.
We consider in Section~\ref{secDiscussion} the possible contribution from several other sources, and we finally summarize our main results and conclusions in Section~\ref{secConclusions}.

\section{A model of the Virgo cluster}
\label{secModel}

In order to characterize the structure of the Virgo cluster, we use two simple analytical models of the gas density and temperature profiles as a function of radius.
The first is a superposition of $\beta$-models, and the second is based on previous work by \citet{AscasibarDiego08}, where an alternative description of the ICM was proposed.  
In both cases, it is assumed that the Virgo cluster is centered at the location of M87, at a distance $d=16.1$~Mpc \citep{Tonry+01}.
At this distance, 1~arcmin corresponds to 4.7~kpc.
Once the physical state of the intracluster gas is specified, the expected X-ray emission and the SZ effect can be computed by evaluating the corresponding integrals along the line of sight.

\subsection{Density and temperature profiles}

The first model we consider is based on the results of \citet[hereafter G04]{Ghizzardi+04}.
These authors applied a spectral deprojection technique to infer the density and temperature distribution of Virgo's central region from a combination of \emph{Chandra}, \emph{XMM-Newton}, and \emph{BeppoSAX} observations. 
They use a superposition of two $\beta$-models for the electron density
\begin{equation}
n_e(r)=
\frac{n_1}{\left[1+(r/r_{1})^2\right]^{\alpha_1}}+
\frac{n_2}{\left[1+(r/r_{2})^2\right]^{\alpha_2}}
\label{eq_beta}
\end{equation}
and an inverted Gaussian for the gas temperature
\begin{equation}
T(r) = T_0 - T_1 \exp\left( -\frac{r^2}{2\sigma_t^2} \right)
\end{equation}
with $T_0 = 2.4$~keV, $T_1 = 0.776$~keV, and $\sigma_t = 20.3$~kpc.


Our second model (which we will refer as AD08 from now on) describes the radial structure of clusters in terms of five free parameters: the total mass $M$ of the system (or the characteristic temperature $T_0$), a characteristic radius $a$, the cooling radius in units of $a$, $0<\alpha<1$, the central temperature in units of $T_0$, $0<t<1$, and the asymptotic baryon fraction at large radii, $f\sim1$.

A thorough description can be found in \citet[AD08 hereafter]{AscasibarDiego08}, where it is shown that this model is able to reproduce the observed X-ray properties of real clusters.
In a few words, the total (gas plus dark matter) density follows a \citet{Hernquist90} profile,
\begin{equation}
\rho(r)=\frac{M}{2\pi a^3}\frac{1}{r/a(1+r/a)^3},
\label{eqRhoH}
\end{equation}
whereas the gas temperature varies according to
\begin{equation}
T(r)=\frac{T_0}{1+r/a}~\frac{t+r/a_{\rm c}}{1+r/a_{\rm c}}.
\label{eqT}
\end{equation}
Hydrostatic equilibrium imposes the mass-temperature relation
\begin{equation}
(n+1)\frac{kT_0}{\mu\Mp}=\frac{GM}{a}
\label{eqMT}
\end{equation}
and the gas density profile
\begin{equation}
\frac{\rho_{\rm gas}(r)}{\rho_0} =
  \left( \frac{1+r/a}{t\alpha+r/a} \right)^{1+\frac{\alpha-t\alpha}{1-t\alpha}(n+1)}
  \frac{\alpha+r/a}{\left(1+r/a\right)^{n+1}}
\label{eqRho}
\end{equation}
where $n$ is an effective polytropic index, $k$ is the Boltzmann constant, $\Mp$ denotes the proton mass, $\mu\simeq0.6$ is the molecular weight of the gas, and $\rho_0$ is the central gas density.
In terms of the cosmic baryon fraction,
\begin{equation}
\rho_0 = f \frac{\Omega_{\rm b}}{\Omega_{\rm m}} \frac{M}{2\pi a^3}
\label{eqFb}
\end{equation}
with $f\sim1$.
We adopt the value $\Omega_{\rm b}/\Omega_{\rm m} = 0.02/0.13 \simeq 0.15$ \citep{Spergel+07}.
As in AD08, we set $n=4$ in order to obtain a constant baryon fraction at large radii.

The two models considered can be seen as a lower and upper limit to Virgo. G04 focuses 
more in the central region of Virgo and does not constrain the most distant regions which can be of relevance for the SZ effect. AD08 on the other hand might be a better description for these distant regions and pusses the temperature to the upper limit boosting the SZ signal while keeping the X-ray emission within the ROSAT constraints. 


\subsection{X-ray luminosity}

For any given density and temperature distributions, the total bolometric X-ray luminosity of the cluster can be computed as the integral
\begin{equation}
L_{\rm X} = \int 4\pi r^2 \epsilon(r)\ \dd r
\end{equation}
of the local emissivity $\epsilon$ over the entire volume of the object.
The X-ray surface brightness at a certain angular position $\vec{\theta}$ would be given by the integral along the line of sight
\begin{equation}
S_{\rm X}(\vec{\theta}) = \frac{1}{4\pi} \int \epsilon(r)\ \dd l
\end{equation}
with $r^2={d^2+l^2-2ld\cos\theta}$, and the total flux would be simply
\begin{equation}
\phi_{\rm X} = \int_{\Delta\Omega} S_{\rm X}(\vec{\theta})\ \dd\Omega
\approx \frac{L_{\rm X}}{4\pi d^2}.
\end{equation}

For thermal bremsstrahlung emission,
\begin{equation}
\epsilon =
\frac{64\pi}{3h}\left(\frac{\pi}{6}\right)^{\!\frac{1}{2}}
\left(\frac{e^2}{4\pi\epsilon_0}\right)^{\!\!3}
\left[\frac{kT}{(m_{\rm e}c^2)^3}\right]^{\!\frac{1}{2}}
\bar{g}\,n_{\rm e}\sum_i{Z_i^2n_i},
\end{equation}
where $h$ denotes the Planck constant, $kT$ is the plasma temperature, $\bar{g}\sim1$ is the average Gaunt factor, $n_{\rm e}$ is the electron number density, and the sum runs over all ionic species.
If one is interested in the energy band between $E_{\rm min}$ and $E_{\rm max}$ rather than the bolometric luminosity, the correction factor $F_{\rm b} = exp(-E_{\rm min}/kT) - exp(-E_{\rm max}/kT)$ should be applied.
In a pure Hydrogen plasma,
\begin{equation}
\frac{\epsilon}{\rm  2.3\times 10^{26}\ erg\ s^{-1}\ kpc^{-3}} = F_{\rm b}
\left(\! \frac{\rho}{\rm M_\odot\ kpc^{-3}} \!\right)^2
\left( \frac{kT}{\rm keV} \right)^{\frac{1}{2}}
\label{eqBrems}
\end{equation}

In real clusters, however, equation~(\ref{eqBrems}) provides a poor approximation at low temperatures, especially when the presence of metals is taken into account.
To make a more rigorous comparison with observational data, we use a MEKAL model \citep{Mewe+85} with abundances from \citet{AndersGrevesse89} to estimate the cooling function of the gas, assuming a constant ICM metallicity of 0.3~Z$_\odot$.
We account for the effective area of the ROSAT PSPC camera by applying the publicly-available response function of the instrument.
The HI column density in the direction of Virgo is small (about $1.6\times10^{20}$ cm$^{-2}$) and does not imply significant absorption.

\subsection{Sunyaev-Zel'dovich effect}

The SZ effect occurs when the free electrons of the intracluster medium interact with CMB photons.
During this interaction, known as inverse Compton scattering, photons gain energy from the free electrons, creating a distortion in the energy spectrum of the CMB proportional to the density and the temperature of the ionized plasma (thermal SZ effect).
Also, if the cluster is moving with a bulk velocity $v$, this velocity creates an additional Doppler shift in the energy of the CMB photons (kinetic SZ), but this contribution is much smaller than the thermal SZ effect and will not be considered in this work.
If the electrons are relativistic, the energy gain can be sufficiently large to boost the CMB photons into the X-ray regime.

In the non-relativistic case, the net (thermal) distortion in a given direction can be quantified by the cluster Comptonization parameter, $y_c$, defined as 
\begin{equation}
y_c(\vec{\theta}) = \frac{\sigma_T k}{m_ec^2}\int T(l)\ n_e(l)\ \dd l .
\label{y_c2}
\end{equation}
The integral is performed along the line of sight through the cluster.
$T$ and $n_e$ are the intracluster electron temperature and density, respectively; $\sigma_T$ is the Thomson cross-section (appropriate at these low energies), $k$ is the Boltzmann constant, and $m_e$ is the electron mass.

The flux and the temperature distortion are more widely used than the Compton parameter because these quantities can be directly measured in observations at mm frequencies.
The distortion in the background temperature is given by
\begin{equation}
 \frac{\delta T}{T} = g(x)y_c  \ ,
\end{equation}
where $x=h\nu /kT_{\rm CMB}$ is the frequency in dimensionless units and $g(x) = x \coth(x/2) - 4$ is the spectral shape factor.
There is no redshift dependence on the temperature distortion, i.e the same cluster will induce the same distortion in the CMB temperature, independently of the cluster distance (except for relativistic corrections).
Regarding the flux, however, there is a strong redshift dependence due to the change of the the apparent size of the cluster with redshift.
The total SZ flux is given by the integral
\begin{equation}
 S_{\rm SZ}(\nu) = \int_{\Delta\Omega} \Delta I(\nu,\vec{\theta})\ \dd\Omega ,
 \label{S_total}
\end{equation}
over the solid angle $\Delta\Omega$ subtended by the cluster. 
$\Delta I(\nu,\vec{\theta})$ is the change in intensity induced by the SZ effect and is given by  $\Delta I(\nu,\vec{\theta}) = I_0f(x)y_c$, where $I_0 = (2h/c^2)(k T_{\rm CMB}/h)^3$ and
\begin{equation}
f(x) =  \frac{x^4e^x}{(e^x-1)^2} \left [ x\,\frac{e^x+1}{e^x-1} - 4\right ] .
\label{fx}
\end{equation}

\begin{figure*}
\centerline{ \includegraphics[width=17cm]{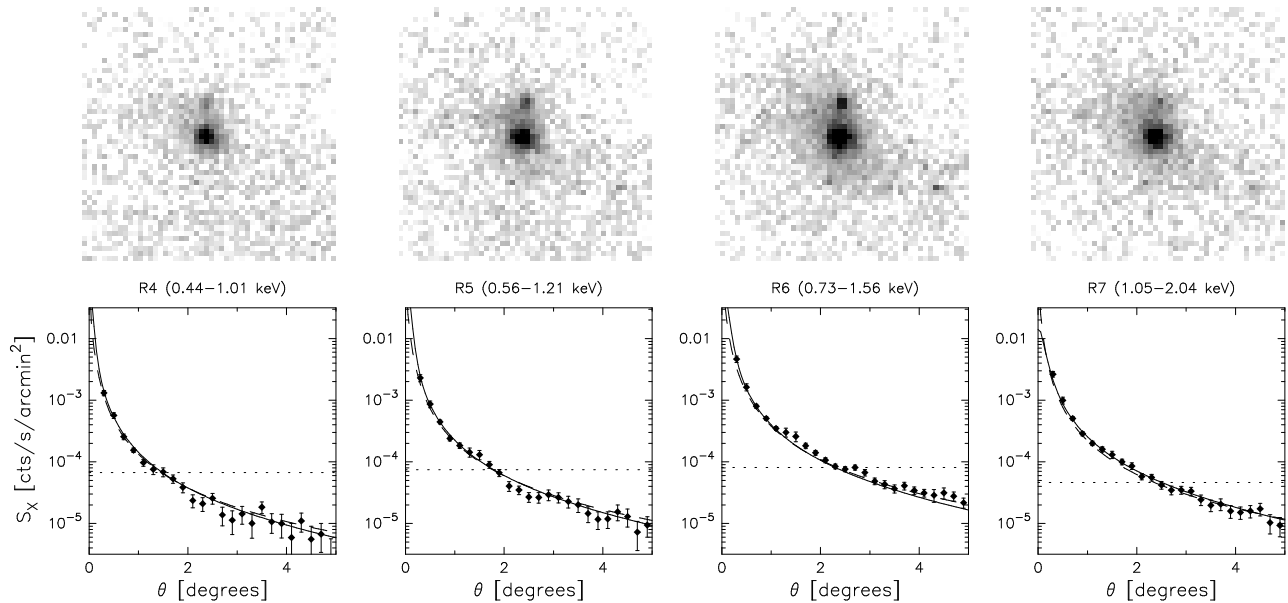} }
\caption{
  ROSAT All Sky Survey images (top) of the Virgo cluster in the X-ray bands R4 (left) to R7 (right).
  All images are $10\times10$ degrees size.
  Bottom panels show the radial X-ray surface brightness profiles around M87.
  Points with error bars represent the foreground-subtracted ROSAT data.
  Solid lines correspond to the AD08 model, and a horizontal dotted line indicates the estimated diffuse foreground in each band (see text for details).
  Dashed lines represent the model by \citet{Ghizzardi+04}.
}  
\label{fig_Xray}  
\end{figure*}

Equation~(\ref{S_total}) can be easily integrated if one assumes that the cluster is isothermal (although in this paper we will solve it numerically).
Then the integral is proportional to $\int\dd\Omega\int n_e(\vec{\theta})\dd l$, which can be reduced to $D_a^{-2}(z)\int dV n_e(\vec{\theta}) \simeq  D_a^{-2}(z) \frac{M_{\rm gas}}{\Mp} =  D_a^{-2}(z) \frac{f_b}{\Mp}M$, where $M$, $f_b$ and $\Mp$ denote the total mass, baryon fraction, and the proton mass, respectively, and the ICM gas has been approximated as a fully ionized Hydrogen plasma.
Finally, one gets
\begin{equation}
 S_{\rm SZ}(\nu) = \frac{3.781 f(x) f_b T M_{15}}{D_a^2(z)} .
 \label{S_total2}
\end{equation}
with the temperature $T$ given in Kelvin, the mass $M_{15}$ in $10^{15}$~M$_{\odot}$, and the angular distance $D_a(z)$ in Mpc. 
In these units, the flux is given in mJy 
(1 mJy = $1.0\times 10^{-26}$ erg s$^{-1}$ Hz$^{-1}$ cm$^{-2}$).

\section{X-ray observations and model fit}
\label{secXray}

We use available X-ray observations to constrain the free parameters of the theoretical models defined above.
We first describe the data and the analysis procedure.
Then, we summarize the best-fitting model of the Virgo cluster and compare it with the results of previous independent studies.

\subsection{Observational data}

Notwithstanding its large integrated X-ray flux, which made possible its detection by the \emph{Einstein} \citep{FabricantGorenstein83} and \emph{Ginga} \citep{Koyama+91} satellites, the large angular extent of the Virgo cluster, about $6^\circ$ in radius, prevented the elaboration of high-resolution maps until the data from the ROSAT All Sky Survey (RASS) became available.
The X-ray surface brightness was found to be strongly peaked at the location of M87, and the large-scale morphology of the ICM traces the galaxy distribution, with three major sub-clusters centered around M87, M86, and M49 \citep{Bohringer+94,Schindler+99}.

The Virgo cluster has also been observed by more recent X-ray observatories, but these instruments feature a narrower field of view, and the new data is restricted to very small regions within the cluster.
Since we are interested on much larger scales, our work is based on the RASS maps of the diffuse X-ray emission \citep{Snowden+97}.
Point sources have been removed from these data, but there is still a strong background of diffuse galactic emission, non-resolved extragalactic AGNs, and local X-ray structures (e.g. the so-called local bubble).
These additional components render the signal from Virgo barely visible at soft energies, so we will only consider the ROSAT bands R4 to R7 (from 0.44 to 2.04~keV) in the present analysis.

For each band, we extracted a patch of the sky of $10^{\circ}\times 10^{\circ}$ size ($50\times50$ pixels) centered in M87.
Virgo is a prominent extended source in all of the surface brightness maps, shown on the top panels of Figure~\ref{fig_Xray}.
The second brightest feature is M86, which can also be clearly distinguished in all bands, most notably R6.

We estimate the contamination by diffuse emission (mostly galactic foreground) and the associated error by computing the median value of the X-ray flux and the semi-difference between the third and the first quartiles\footnote
{
Such procedure is equivalent to computing the mean and the standard deviation for a Gaussian distribution, but it is substantially more robust to the presence of outliers in the data set.
} in maps of $16^{\circ}$ (80 pixels) side, iteratively removing the highest intensity points until they are within three sigma of the estimated uniform background.
The bottom panels plot the surface brightness profile in each band.
The estimated foreground emission ($67$, $75$, $81$, and $46$~counts~s$^{-1}$~arcmin$^{-1}$ for R4, R5, R6, and R7, respectively) is indicated by the dotted lines.


\begin{figure}
\centerline{ \includegraphics[width=8cm]{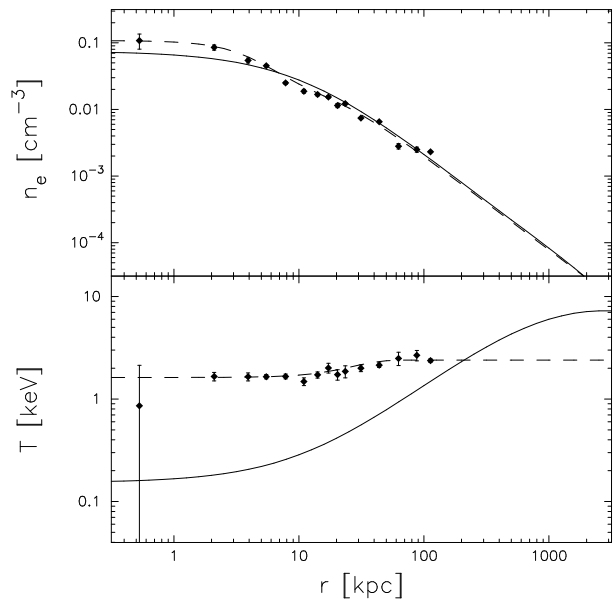} }
\caption{
  Three-dimensional electron density (top) and temperature (bottom) profiles around M87.
  Data points correspond to the results of G04, and dashed lines represent the best fit provided by these authors.
  The AD08 model is shown by the solid lines.
}  
\label{fig_Profiles}  
\end{figure}
\subsection{Results}


The X-ray emission is used to set the values of the AD08 model parameters.
We employ a particle swarm optimization algorithm to fit the ROSAT data in the four bands simultaneously.
The minimum chi-square (51.37 for 55 degrees of freedom) is obtained for $M=7.3114\times10^{15}$~M$_{\odot}$, $a=3.5786$~Mpc, $\alpha=0.1112$, $t=0.0177$ and $f=0.4995$.

The corresponding density profile, plotted on the top panel of Figure~\ref{fig_Profiles}, is very similar to the results of G04.
This is quite remarkable when one takes into account that the resolution of the ROSAT maps is only $56$~kpc.
The temperature structure, shown on the bottom panel, is however significantly different.
The AD08 model predicts a lower temperature in the inner regions and a hotter ICM in the outer parts.
The discrepancy near the center is not very relevant for the present study.
This region is very small and does not contribute much to the integrated X-ray flux nor the SZ effect.
Moreover, it will be strongly masked out due to the presence of point sources in WMAP data (see below), and it is likely that some contamination from the central galaxy affects the X-rays as well.
At large distances, G04 assume a constant temperature of $2.4$~keV, while the ICM gas in the AD08 model reaches values up to $\sim 7$ keV and (see figure \ref{fig_Profiles}) then drops towards larger radii.
Although the X-ray emission is almost identical in both models (see Figure~\ref{fig_Xray}), it will be shown that the temperature profile plays a crucial role in the SZ effect.

\section{WMAP observations and Sunyaev-Zel'dovich effect signal}
\label{secWMAP}

Given a model of the ICM, it is straightforward to predict the expected SZ signal, but the comparison with WMAP data is far from simple.
As opposed to X-rays, it is not possible to see Virgo directly in WMAP because the SZ effect is weaker than the underlying CMB.
We discuss now the pre-processing of the mm data as well as the statistical basis of the comparison between models and observations; then, the outcome of such a comparison and an assessment of its significance.

\subsection{Observational data}

We consider a region of $25.65^\circ\times25.65^\circ$ around the location of M87, extracted from the WMAP-3yr data set \citep[see for instance][]{Hinshaw+07}.
To reduce the CMB contamination, we perform a basic cleaning process that makes use of the well-known frequency independence of the CMB anisotropies.
We consider the foreground-reduced maps in the Q, V, and W bands, corresponding to frequencies of 41, 61, and 94 GHz, respectively.
The Foreground Template Model discussed in \citet{Hinshaw+07} and \citet{Page+07} has been subtracted from all the unreduced sky maps.

A simple linear combination of the WMAP channels can produce a new map where the CMB vanishes, at the expense of removing a significant fraction of the SZ signal and increasing the noise level.
Also, the resolution of the V and W channels must be degraded to match that of the Q channel.
Our choice, V+W-2Q, tries to maximize the SZ residual.
Due to the frequency dependence of the SZ effect, this combination has positive sign in those areas where the SZ effect is dominant with respect to other sources.
From now on we will refer to the linear combination V+W-2Q as the \emph{residual SZ map}.

Unfortunately, the CMB is not the only contaminant that must be dealt with.
As will be discussed in more detail in Section~\ref{secDiscussion}, there is also contamination by point sources and diffuse emission of different kinds.
We removed the monopole and the dipole in the residual (all sky) SZ map to correct for possible calibration differences between the different channels, and we used the mask provided by WMAP to remove the brightest point sources in the field of view.
There are 4 of these sources near the center of Virgo and 13 more in the rest of the field of view.
Masking them reduces again the strength of SZ signal, especially in the central part, where it should be maximum.
However, most point sources feature a synchrotron-like spectrum, and therefore they yield a negative contribution to our residual SZ map.
If we did not apply the mask, we could even find a net negative signal at the position of Virgo.

\begin{figure}
\centerline{ \includegraphics[width=8cm]{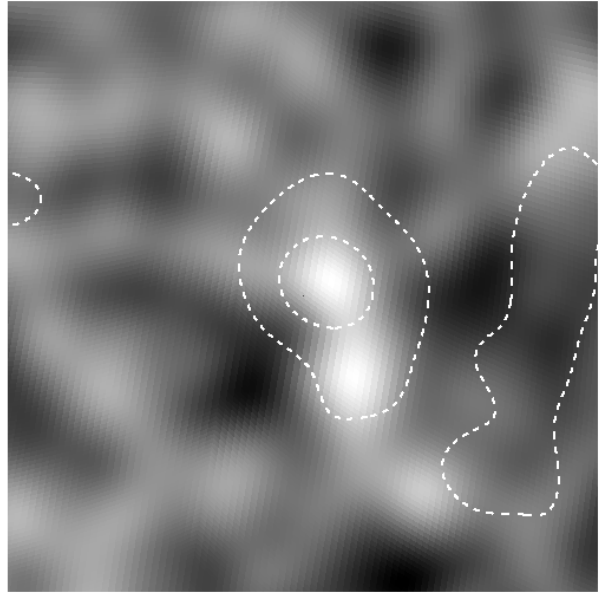} }
\caption{3 degree smoothed WMAP residual ($V+W-2Q$). Contours show the Virgo 
            region (smoothed with a 3 degree fwhm Gaussian) 
            as seen by ROSAT in the R6 band ($0.9-1.3$ keV)
            The structure to the right of Virgo 
            in ROSAT data is diffuse emission from the Galaxy. The contour levels correspond to 
            $115\times 10^{-6}$ cts s$^{-1}$ arcmin$^{-2}$ and $300\times 10^{-6}$ cts 
            s$^{-1}$ arcmin$^{-2}$.
            In this and the other plots, the field of view is 25.65 degrees 
            on a side and centered on Virgo.}
\label{fig_ROSAT_WMAP_Virgo}
\end{figure}
At this stage, the map still seems at first sight to be completely dominated by instrumental noise.
However, this component can be well described by white Gaussian noise, and therefore its magnitude can be reduced by averaging over a large area.
As long as we do not exceed the scale of the cluster, the SZ signal from Virgo will be enhanced by this process, because the amplitude of the noise (the dispersion around the average, which is 0) decreases as the square root of the number of pixels involved, or, equivalently, the characteristic length of the smoothing kernel.
The result obtained after excluding point sources and filtering out the noise with a Gaussian of FWHM$=3^{\circ}$ can be seen in Figure~\ref{fig_ROSAT_WMAP_Virgo}.
There is a positive fluctuation (as expected for a SZ signal) at the center.
This signal seems to extend a few degrees south, apparently correlated with the X-ray emission.
It could be a genuine SZ signature, but also a fluctuation of the noisy background 
or even a contribution from unresolved sources with a SZ-like spectrum.

Interestingly, the mean intensity of the residual SZ map deviates significantly from zero.
Such deviation is neither compatible with the expected mean SZ signal from Virgo nor with a large scale fluctuation of the Gaussian noise.
We will discuss more about this issue later, but we can anticipate that the data suggests the presence in this part of the sky of a diffuse component, possibly of galactic origin, not related to the SZ effect.

\subsection{Statistical analysis}

A mere visual inspection\footnote{A technique that usually rivals, and often surpasses, most sophisticated data analysis algorithms.} of Figure~\ref{fig_ROSAT_WMAP_Virgo} indicates that there is indeed a positive fluctuation in the residual SZ map at the position of the Virgo cluster.
It is our goal to find out, \emph{quantitatively},
\begin{enumerate}
\item whether the signal constitutes a statistically significant detection, or it can be understood as a random fluctuation of the instrumental noise.
\item In either case, whether it is compatible or not with the expectation from the models based on X-ray data.
\end{enumerate}

We can address both questions by computing the probability that the measured intensity arises from pure Gaussian random noise, and compare it with other process in which the average equals the expected SZ signal.
To compute these probabilities, we carried out Monte Carlo simulations based on our analytical model, G04, and the null hypothesis.
All of them involve the addition of fake instrumental noise, application of the point source mask, and Gaussian smoothing, mimicking the analysis of the real WMAP data.

We account for the random noise by generating 400 realizations using the published sensitivity per channel and the number of observations per pixel.
A realization of the noise is made for each of the 8 channels involved (Q1,Q2,V1,V2,W1,W2,W3, and W4).
These channels are combined to produce the Q, V, and W maps.
The resulting V and W maps are then degraded to the resolution of Q, and the linear combination V+W-2Q is constructed.

\begin{figure}
\centerline{ \includegraphics[width=8cm]{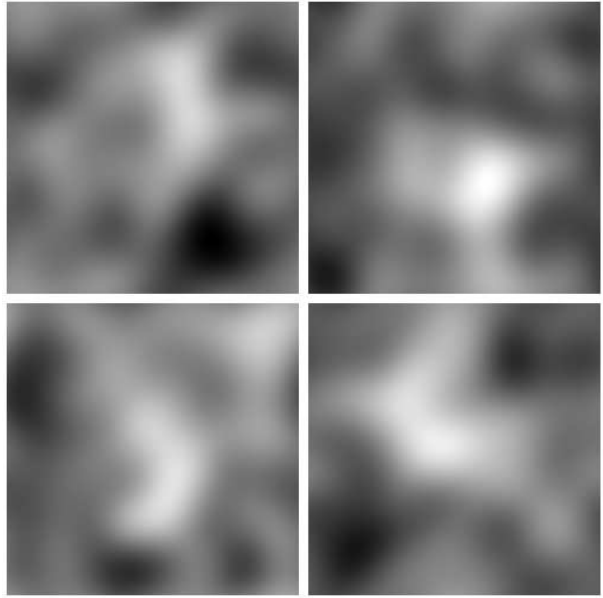} }
\caption{5 degree smoothed simulated WMAP residual maps ($V+W-2Q$) centered 
  on Virgo (AD best fitting model). The maps include simulated WMAP noise and Virgo SZ 
  contribution. Each map corresponds to a different realization of the 
  noise.
}
\label{fig_4Sims}
\end{figure}

To these simulated noise maps we add the predicted SZ effect from Virgo, according to our two different models.
The WMAP point source mask is then applied, and the map is convolved with a Gaussian filter of 5 degrees FWHM.
This scale roughly corresponds to the size of the Virgo cluster, and it is nearly optimal in the sense that it integrates most of the signal and filters out the random noise component.
We have tried with other filters (Mexican hat wavelet, matched filter) with similar results. We have also tried other scales (see table \ref{tabSignal}) but 5 degrees seem to 
render optimal results.

\begin{table}
\caption
{Amplitudes (in $\mu$K) of the data ($V+W-2Q$) at the Virgo position and at different filtering 
scales compared with the dispersion of the data and simulated noise at those scales. An average signal of 7 $\mu$K has been subtracted and a 1 $\mu$K correction has been added to the data (see text). The scales represent the FWHM of a Gaussian in degrees.}
\begin{tabular}{lcccc}
\hline
Scale          & $3^{\circ}$ & $5^{\circ}$ & $7^{\circ}$ & $10^{\circ}$ \\
WMAP Virgo     & 22.2 & 10.4  & 6.2 & 3.6 \\
WMAP $\sigma$  & 6.1  & 3.5  & 2.3 & 1.4 \\
Noise $\sigma$ & 5.7  & 3.3  & 2.4 & 1.6 \\
\hline
\end{tabular}
\label{tabSignal}
\end{table}

Figure~\ref{fig_4Sims} shows some examples of simulated noise+Virgo+mask maps after combining the V+W-2Q channels and applying the Gaussian filter.
The 5 degree Gaussian integrates the signal over a large area, giving a better significance at the center, but it misses some interesting structures, like the brighter spot south of the central region.
Comparing figures~\ref{fig_ROSAT_WMAP_Virgo} and~\ref{fig_4Sims}, though, we see that the non-symmetric features observed in WMAP can be reproduced in simulations that include realistic instrumental noise in spite of the intrinsic symmetry of the models.

\subsection{Results}
By performing hundreds of Monte Carlo simulations, we obtained an estimate of the average expected signal for each model as well as the typical variation due to the Gaussian noise fluctuations.
The variance of the filtered noise, based on our 400 realizations, is $\sigma_{\rm noise}= 3.3 \mu$K on scales of 5 degrees (see Table~\ref{tabSignal}).
The same table shows the observed signal and dispersion of the data and simulated noise at different scales.
It is important to note that the mean intensity over the whole field of view, $\left<{\rm V+W-2Q}\right>=7~\mu$K (see the discussion in Section~\ref{secDiffuse} below), has been subtracted from the values quoted in the table. This mean signal is associated to an unidentified extended background in this region of the sky. To account for the expected signal from Virgo we add 
the corresponding average signal expected from our model (in the considered field of view) which is 1 $\mu$K. Hence, the observed field of view has an average signal of 1 $\mu$K. 

Comparing the WMAP values to the variance of the noise on different scales, one sees that the observed intensity of the residual SZ map would represent a 3-$\sigma$ fluctuation for a FWHM of 3 degrees.
Though unlikely, it would not be completely impossible that the signal arose from a random fluctuation.
The probability of such an event, the $p$-value, is of the order of 1 per cent.
The evidence against the null hypothesis (i.e. no signal, just noise) becomes even weaker as we increase the smoothing scale.
On scales of $10^\circ$, the observed fluctuation is only $2.5$ sigmas above the noise level. Comparing the two models with the observed signal, we predict on average a 1.4-$\sigma$ signal for the G04 model and a 2.1-$\sigma$ signal for the AD08 model  on scales of 5 degrees (see table \ref{tabSignal2}). This signal is lower than the observed one but still compatible with it as we will see.

A complementary analysis can be accomplished by studying the location and intensity of the maxima of the simulated model plus noise data.
If the observed peak was actually due to a random fluctuation, there is no reason why it should coincide approximately with the center of Virgo.

\begin{table}
\caption
{
Predicted amplitudes (in mK) of the different models after a 5 degree Gaussian 
smoothing, evaluated at the position of Virgo ($\sigma_V$) and at the point where 
the total intensity (signal+noise) is maximum ($\sigma_M$). 
We also show the dispersions at the position of Virgo and at the maximum. 
The value of $\sigma_V$ for WMAP data corresponds to the dispersion of the signal 
in the Virgo region.
Since WMAP provides us with only one data point the value of $\sigma_M$ for 
WMAP can not be determined. 
}
\begin{tabular}{lcccccc}
\hline
& WMAP & G04 & AD & No SZ \\
\hline

Virgo         & 0.0102  &  0.0045   &  0.0070   &  0.0    \\
Maximum       & 0.0105  &  0.0090   &  0.0107   &  0.0076 \\
$\sigma_V$    & 0.0034  &  0.0033   &  0.0033   &  0.0033 \\
$\sigma_M$    &   -     &  0.0018   &  0.0023   &  0.0018 \\
\hline
\end{tabular}
\label{tabSignal2}
\end{table}

\begin{figure}
\centerline{ \includegraphics[width=8cm]{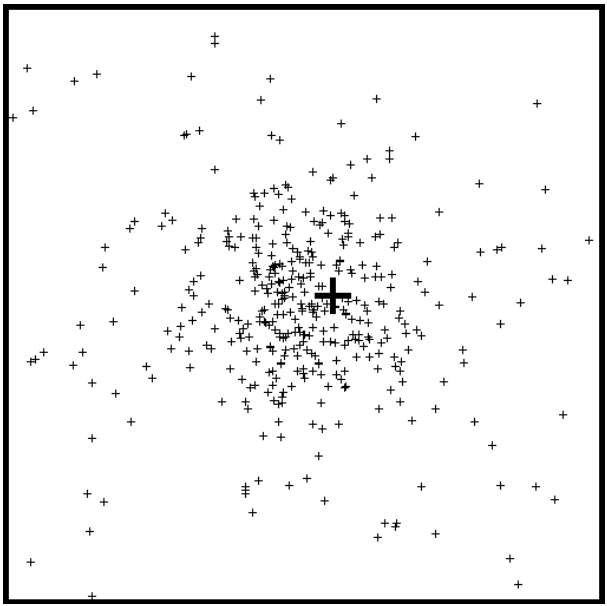} }
\caption{Positions of the maxima in the 400 simulations for model AD08. 
  The position of WMAP data maxima is marked with a big cross.
}
\label{fig_Maxima}  
\end{figure}
Two kinds of analysis were made based on the maxima. Again, we choose the 5 degree Gaussian 
to perform the smoothing. The first analysis looks at the distribution 
of the maxima in the field of view. The result can be seen in figure \ref{fig_Maxima}.
The maxima distribute around the center of the image but there is a large dispersion of several degrees around it.
The position of the maxima of WMAP is close to the center.

\begin{figure}
\centerline{ \includegraphics[width=8cm]{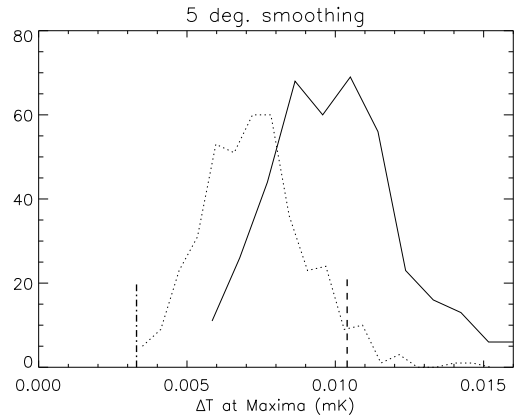} }
\caption{The histograms show the frequency of the temperatures at the maxima 
  for model AD (solid line) and for a simulation with just noise (dotted line). 
  The vertical dot-dashed line corresponds to the dispersion of the 
  background noise and the dashed line is the observed value of WMAP data at 
  the position of the maximum.
}  
\label{fig_Histo2}
\end{figure}
The second analysis looks at 
the values taken at the maxima by the simulated data. A histogram of these values 
is shown in figure \ref{fig_Histo2}.
In the same figure we also show the histogram of the maxima 
if only the noise is included in the simulation. From the previous results the main conclusion 
is that the observed WMAP fluctuation around Virgo is compatible with the expected one from 
simulations which are compatible with the X-ray constraints. The results based on the maxima might be more 
conclusive than the previous ones (where we looked at the values at the centre of Virgo). 
This is emphasized also in table \ref{tabSignal2} where the average of the simulated maxima 
coincides almost perfectly with the observed one. 

The main conclusion is that the observed WMAP signal around Virgo could
be a $3\sigma$ random fluctuation, but it is far more consistent with the
prediction of models of the SZ effect compatible with the X-ray constraints.
The results based on the maximum strengthen this conclusion. As can be seen
in table \ref{tabSignal2}, the average intensity of the simulated maxima coincides.

\section{Discussion}
\label{secDiscussion}
We have shown how observed signal is consistent with 
the predictions made by models which are compatible with the observed X-ray emission 
in Virgo. In the previous section, we only added instrumental noise to the models of
the SZ effect. However, other sources might contribute significantly to the
observed signal. Here we consider contamination by unresolved point sources,
as well as diffuse emission from the warm-hot intergalactic medium and other
residual foregrounds. 

\subsection{Point sources}

Point sources can change the amount of SZ signal inferred from WMAP data. In the case of Virgo this is even more true. For instance, the brightest source in the sky in the Ka WMAP band is precisely M87 at the centre 
of the Virgo cluster (with more than 19 Jy in Ka). M87 is known to be one of the brightest 
radio sources in the Universe. Other bright radio sources can be found near M87. 
If we do not apply the WMAP mask, these bright point sources near the centre produce a negative 
(instead of positive) signal in the combination $V+W-2Q$. 
After masking these sources, a residual point source 
signal can still be left in Virgo that reduces the significance of the SZ effect. It is 
hard to evaluate this component as we ignore the number and intensity of such sources 
in Virgo. Assuming the sources are randomly distributed over the field of view simplifies  
the estimation of the point source contribution. 
We have made simulations of the contribution of field point sources 
below 0.5 Jy (which was the threshold used by WMAP to mask point sources).

\begin{figure}  
   \centerline{ \includegraphics[width=8cm]{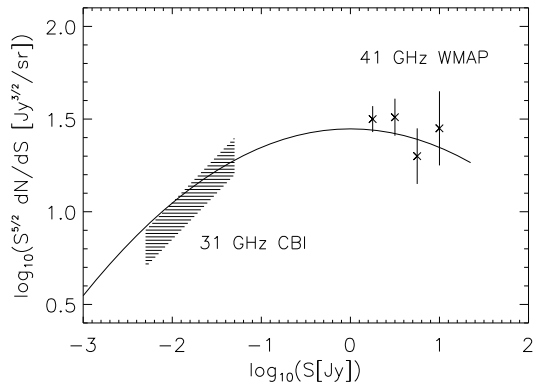} }
   \caption{The solid line shows the number counts of the model used to simulate 
            the point source population under 0.5 Jy at 41 GHz. WMAP counts in  
            this band are shown together with the error bars. Also CBI counts at 
            31 GHz. 
           }  
   \label{fig_PS}  
\end{figure}
  
In figure \ref{fig_PS} we show the model used to make the simulations at 41 GHz 
(Q band in WMAP). This model is compatible with the number counts derived 
in the 30-45 GHz range from CBI, VSA, DASI and WMAP Q band 
(see figure 13 in Bennet et al. 2003). Our simulations predict a mean signal 
from point sources ($S<0.5$ Jy) in our field of view of $8.8 \mu K$ and a 
dispersion of $8.5 \mu K$ in WMAP Q band. The dispersion reduces to $0.75 \mu K$ when 
the point source field is smoothed with a 5 degree fwhm Gaussian (see discussion 
below).  
The dispersion is compatible with the predictions of Toffolatti et al. (1998). 
There is little information about the point source populations at higher WMAP 
frequencies (V and W bands).  Assuming the predictions from 
Toffolatti et al. (1998) the dispersion from point sources in WMAP V band should 
be  $\sim 6 \mu K$ and in WMAP W band of  $\sim 2 \mu K$. Therefore the non-removed 
sources are not expected to contribute significantly to the signal if they are 
distributed uniformly (note that these dispersions will be reduced to less than 1  
$\mu K$ when filtered with the 5 degree fwhm Gaussian). 
 
\begin{figure}  
   \centerline{ \includegraphics[width=8cm]{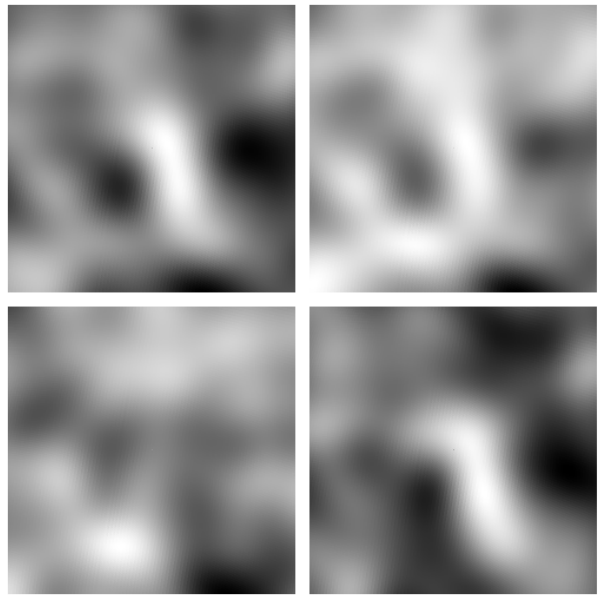} }
   \caption{5 degree smoothed WMAP residual of the four differences 
            $V+W-2Q$ (top left), $W-Q$ (top right), $W-V$ (bottom left) and 
            $V-Q$ (bottom right). 
           }  
   \label{fig_WMAP_Virgo4}  
\end{figure}  

\begin{figure}  
   \centerline{\includegraphics[width=8cm]{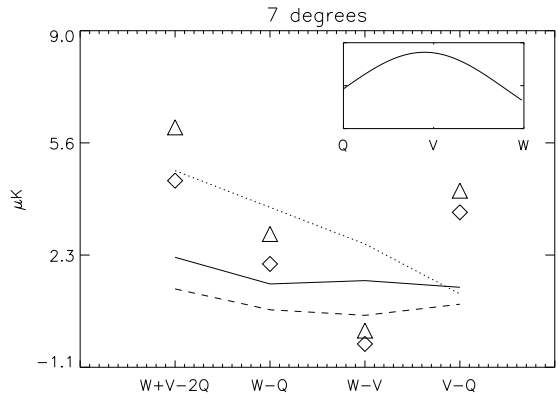} }
   \caption{Observed versus predicted signal for the different combination of 
            channels after masking the brightest sources and smoothing with a 
            7 degree FWHM Gaussian. The symbols (triangles) show the observed 
            signal in the different combinations. The diamonds is the same result 
            but using the uncleaned WMAP data channels. 
            The dotted line shows the predicted SZ signal (mask included). 
            Dashed line shows the predicted RMS of the 
            background of radio sources with $S < 0.5 Jy$ at 44 GHz and the solid line 
            shows the expected RMS of the instrumental noise. The inner plot shows the 
            suggested spectrum of the extragalactic sources inside Virgo.   
           }  
   \label{fig_Sigma_Signals}  
\end{figure}

\begin{figure}  
   \centerline{\includegraphics[width=8cm]{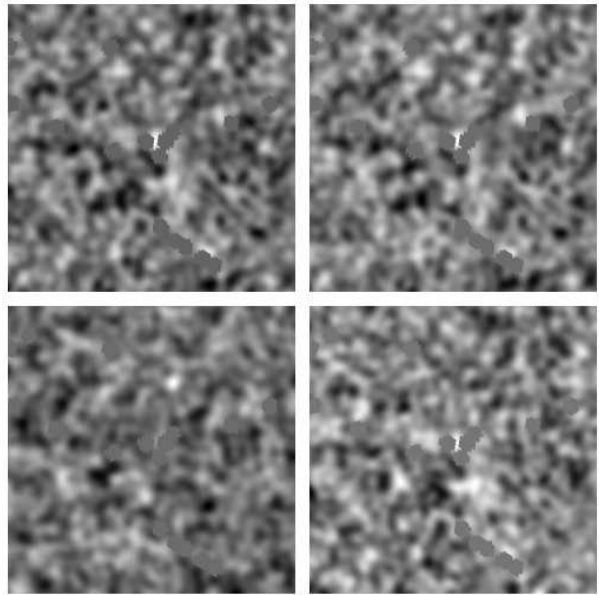} }
   \caption{Difference maps filtered with 1 degree fwhm Gaussian to smooth the data. 
            The WMAP mask is shown as well (grey pixels). Top left $V+W-2Q$, top right $W-Q$, 
            bottom left $W-V$ and bottom right $V-Q$.
            At the centre of Virgo it can be appreciated what it seems like a non-removed source with 
            a inverted spectrum peaking somewhere between the V and W bands. 
            Other potential sources can be 
            seen for instance to the north of Virgo's centre in the $W-V$ map (bottom left). 
           }  
   \label{fig_WMAP_4diffs_1deg}  
\end{figure}

Our results show that the effect of uniformly distributed field sources with 
fluxes under 0.5 Jy can be ignored as they do not change significantly the Virgo 
flux in $V+W-2Q$. About the point sources within Virgo with less than 0.5 Jy, 
one could expect a total contribution of several Jy. This is less than the expected 
SZ flux of Virgo at WMAP frequencies which due to its small distance to us can be 
as much as 50 Jy. We test the source contribution from non-resolved Virgo sources by 
repeating the simulation exercise above (in the Q band) but this time we introduce 
a 5 percent overdensity of sources (with less than 0.5 Jy) on a $5\times5$ degree box 
centered on Virgo and filter the field with a 5 degree fwhm Gaussian. The 5 percent 
excess of sources predicts about 170 sources with fluxes between 5 mJy and 0.5 Jy in 
the $5\times5$ degree area. This estimate is uncertain as we do not know how many 
radio sources are within the Virgo area but we can consider this estimate as a 
conservative one. The resulting source field shows an excess of $7 \mu$K (in Q band) at the Virgo 
position with respect to the mean temperature (the dispersion is about 1 $\mu$K). 
For comparison, Virgo is expected to show a $25 \mu$K decrement (in Q band) on its peak 
with respect to the mean temperature of the CMB when filtered at the same scale. The 
conclusion that we can extract from this experiment is that unresolved radio sources 
within Virgo may still play a crucial role in contaminating the SZ signal. 
In particular when we are working with the differences of WMAP channels. 
By doing these differences, a significant amount of the SZ signal is thrown away. 
We can predict exactly how much because we know the frequency dependence of the 
SZ effect. With the point sources 
however, the situation is a bit different. Little is known about the population of 
extragalactic sources at WMAP frequencies. There is some evidence that most sources 
at WMAP frequencies can be well characterized by a flat spectrum (Bennet et al. 2003) 
but there is a wide variety of spectral indexes among them. The most striking 
hint about their possible contribution to the WMAP frequency channels comes from 
the data itself. By looking at the individual differences we see that 
the SZ signal in Virgo vanishes in the $W-V$ difference while in $V-Q$ is maximum  
(see figures \ref{fig_WMAP_Virgo4} and \ref{fig_Sigma_Signals}). These differences suggest that the Virgo 
zone might be dominated by a population of galaxies with a spectrum peaking 
in the V band (see small plot in figure \ref{fig_Sigma_Signals}). A spectrum like 
this can explain the observed signal (triangles in figure \ref{fig_Sigma_Signals}) 
as a combination of SZ plus radio sources with a spectrum peaking at 60 GHz 
(in temperature). 
This combination explains why the signal vanishes 
in the $W-V$ combination map (the sources cancel out the SZ signal) while producing 
a strong signal in the $V-Q$ combination map (the sources dominate the weaker SZ signal). 
It also explains the relatively low SZ signal in the $W-Q$ combination map (the 
sources reduce partially the strength of the SZ effect). 
The possibility than non-removed sources play an important role as potential contaminants of the 
SZ signal gains additional support when we look at the difference maps but this time filtered 
with a 1 degree fwhm Gaussian (see figure \ref{fig_WMAP_4diffs_1deg}). 
In this case a bright signal is seen extending from M87 (which shows a strong jet in the 
radio and in the optical) and beyond the WMAP mask. 
The amplitude of this signal is 100 $\mu$K in $V+W-2Q$, 65 $\mu$K in $W-Q$, 33 $\mu$K in $W-V$, 
and 35 $\mu$K in $V-Q$ (the corresponding noise dispersions when filtered with a 1 degree 
fwhm Gaussian in these difference maps are 19, 12, and 10 $\mu$K respectively). 
Since M87 is a very strong radio source one could think that part of the signal lays in the 
sidelobes outside the WMAP mask (this is indeed a realistic possibility). 
This, however, is an unlikely explanation as WMAP clearly determined 
a synchrotron-like spectrum for M87 in $Q$, $V$ and $W$ 
and the frequency dependence of the bright residual near M87 
shows an opposite behavior. This fact suggests that the central bright residual comes 
from an unresolved source near M87 in Virgo or is in fact SZ effect between the 
galaxies M87 and M86 coming from a denser and hotter gas phase. The latter case is supported 
by X-rays  observations where a significant amount of X-ray emission comes from the 
area between these two galaxies. 
If this is true, the lack of signal in the $W-V$ difference map can be understood as 
due to a positive noise fluctuation ($\approx 3\sigma$) in the V map at the position 
of the bright residual. This also explains why the $V-Q$ map 
shows a signal comparable in strength to the $W-Q$ since in this case the positive noise fluctuation would go in the same direction as the SZ signal. 
Masking out the central bright residual, the amplitude in the centre decreases 
about 20\% with respect to the result presented in figure \ref{fig_WMAP_Virgo4}.
 
Finally one could argue that the strange behavior of the signal in the Virgo zone (at large scales) 
or at least part of it is due to poorly removed Galactic components. 
This is however unlikely since the same behavior is seen when the 
uncleaned WMAP data is used instead (diamonds in figure \ref{fig_Sigma_Signals}). See however the discussion below.


\subsection{Diffuse emission}
\label{secDiffuse}

\subsubsection{WHIM around Virgo}

One important application of the SZ effect or X-ray observations is the detection of 
the warm hot intergalactic medium (or WHIM hereafter). 
A large fraction of the baryonic matter is supposed to be in this form and emit (or absorb) 
in the X-rays. The WHIM could also distort the CMB spectrum through Compton scattering. 
So far, the WHIM has eluded a clear detection. Nearby WHIM around Virgo, should produce a SZ 
signal expanding over many degrees while it might escape detection in the X-ray band. 
The WHIM is predicted to be most dense around clusters of galaxies with densities ranging from 
$10^{-6}$ cm$^{-3}$ to $10^{-4}$ cm$^{-3}$ and temperatures between $10^6$ and $10^7$ Kelvin.
Fujimoto et al. (2004) made an attempt to measure the WHIM associated to Virgo using XMM-Newton 
data. Their results suggested the presence of WHIM around Virgo with a mean electron density of 
about $6\times 10^{-5}$ cm$^{-3}$ and a line-of-sight length of 9 Mpc. 
We model a possible WHIM cloud around Virgo as a $\beta$-model with a 1.5 Mpc radius, 
a central (peak) electron density of $10^{-4}$ cm$^{-3}$ and a constant temperature of 2.5 keV.
We also integrate the $\beta$-model over 9 Mpc.
We have checked that this model does not exceed the observed X-ray emission around Virgo and it is 
compatible with the observations predicting an excess (with respect to the models considered above) 
of about $30\times 10^{-6}$ cts s$^{-1}$ arcmin$^{-2}$  at the centre of Virgo and dropping to 
about $3\times 10^{-6}$ cts s$^{-1}$ arcmin$^{-2}$ extra counts at 1.5 Mpc from the centre 
of Virgo (in ROSAT R6 band). 

Regarding the SZ effect, the superposition of this extended component with the models considered 
above increases the measured SZ signal by about 20 percent when filtered with a 7 degree 
fwhm Gaussian. This is not sufficient to arrive to any conclusion about the nature and amount 
of the WHIM around Virgo. It also does not explain the deviation from zero of the mean of the 
signal since this deviation (7 $\mu$K in $V+W-2Q$) is several times larger than the 
predicted by our WHIM model ($\approx 1 \mu$K in $V+W-2Q$).  
Future data, like Planck data, will have the sensitivity to deepen in this 
issue and possibly detect directly the WHIM around Virgo through its interaction with the 
CMB photons.

\subsubsection{Residual galactic or extragalactic foregrounds}
We mentioned earlier how the mean of the $V+W-2Q$ WMAP map is significantly different from zero. 
This deviation from zero can not be explained only by the presence of a SZ component 
or a noise fluctuation (let us recall that the WMAP difference maps used in this 
analysis have been monopole and dipole subtracted). 
To get a sense of the underlying signal we filter the residual SZ map with a scale of 
12 degrees. The mean of our simulated maps (noise plus Virgo) when filtered 
with a scale of $12^{\circ}$ is 1 $\mu$K and with the same dispersion of 1 $\mu$K. 
On the other hand, the mean of the observed region of Virgo is almost one order of 
magnitude larger (7 $\mu$K) which is approximately a 7$\sigma$ deviation.
One possible interpretation of this component is that it is of galactic origin. However, 
a visual comparison of this component with maps of the synchrotron, free-free, 
and dust emission (all of them filtered at the same scale of 12 degrees) shows no obvious
correlation. A combination of these three components could produce a pattern 
like the one observed by WMAP. Another possibility is that it is 
due to spinning dust since this component was not subtracted but this possibility is unlikely because in this case it would 
contribute with a negative signal since it is expected to be stronger in 
the $Q$ channel. 
Finally, as mentioned earlier, residual point sources could be contaminating the 
signal leaving a positive residual in the $V+W-2Q$ map. Estimating their contribution 
on large scales is not a trivial task since it will depend crucially of their clustering. 
As seen above, a 5\% increase on a 5 degree scale in the clustering of sources with fluxes 
less than 0.5 Jy in Q can produce an excess of up to 7 $\mu$K in Q. It is then 
tempting to speculate that these sources could explain the deviation in the mean signal on 
this part of the sky. More work will be needed in the future to explain this signal. 
As discussed before, to account for this {\it mysterious} component, we subtracted its mean 
value (7 $\mu$K) to the observed $V+W-2Q$ map (and compensated the Virgo signal by 
adding 1 $\mu$K which is the expected mean signal for Virgo).


\section{Conclusions}
\label{secConclusions}

Our analysis shows that there is a $3\sigma$ signal in WMAP at the position of Virgo when 
the difference of the data $V+W-2Q$ is filtered on scales of 5 degrees. 
The value of $\sigma$ is estimated from simulations of the instrumental noise where the 
sensitivity per pixel and angular resolution for each channel has been simulated 
following the WMAP specifications. 
Using X-ray data (ROSAT) we build 
spherically symmetric models of the gas which are compatible with the X-ray data. These 
models (plus WMAP noise) predict on average a $1.4\sigma$ (G04) and $2.1\sigma$ (AD08) 
signals (after 5 degree smoothing), i.e compatible with the observed one at the 
$1\sigma$ level. However, the AD08 model reproduces better the observations. 
An analysis looking at the maxima in the simulated maps and the observed data makes the 
agreement even stronger ($3\sigma$ in both cases). 
The simulations reveal that Virgo's SZ signal 
in WMAP data is significantly affected by the instrumental noise, but also that the 
$3\sigma$ signal is possibly due to the SZ effect. 
We discussed the existence of a compact structure  
between the central M87 and the sub-dominant (in X-rays) M86 galaxies. This feature has a frequency 
dependence that is compatible with a SZ-like spectrum (plus a 3$\sigma$ noise fluctuation 
in the V map at that position). The existence of a significant amount of diffuse X-rays between 
M87 and M86 favors this possibility. 
The compact feature is also compatible with an unresolved source between M87 and M86. 
If this feature is masked, the total integrated signal on scales of 5 degrees decreases by 20\% 
at the centre of Virgo. This feature has not been included in our simulations and could increase 
the predicted SZ signal. Estimating how much this increase would be is not trivial since the 
area between M87 and M86 is heavily masked by the WMAP mask.

We also find other surprises when analyzing the data. The mean of the observed data 
in the field of view deviates approximately $7\sigma$ from the expected 
mean when only Virgo's SZ and the instrumental noise are included in the simulations. 
This suggests that other component is present in the data and  with a frequency dependence 
such that the mean in the $V+W-2Q$ difference map is about $7 \mu$K in the considered field 
of view. We consider three  possible candidates, extragalactic sources,  
non-removed (or removed in excess) foreground galactic components and the WHIM. 
The first candidate could explain part of the deviation if the sources are 
clustered in the Virgo area. Since we are looking at a cluster region this is a serious possibility to be considered. To check for the second candidate we inspect visually different foreground maps of this region, but we do not 
find any clear correlation between them and the observed diffuse background in $V+W-2Q$. 
Finally, we considered the WHIM as a possible explanation for the deviation of the mean signal. 
Considering a WHIM model compatible with the X-ray constraints we predict the WHIM to contribute 
with approximately $1 \mu K$ to the mean in $V+W-2Q$. However, this is not sufficient to explain the $7 \mu$K 
deviation. Further work will be needed to clarify these and other issues. In particular, special 
attention needs to be paid to the role of contaminating point sources, features in the gas 
distribution, and other signals coming from our own galaxy or WHIM.

Using the 6-year WMAP data will make possible to reduce $\sigma_{\rm noise}$ by a factor $\sqrt{2}$. The statistical evidence for the peak to be a genuine detection would be more convincing, but perhaps not enough to establish a firm conclusion. On the other hand, the superior capabilities of Planck offer an opportunity to improve significantly the results presented in this paper.

\section{Acknowledgments}  

The authors would like to thank M. Ceballos and E. Brait for useful comments.
This work has been funded by the Spanish \emph{Ministerio de Educaci\'on y Ciencia}, 
project AYA2006-06266, and JMD benefits from a contract under the \emph{Ram\'on y Cajal} 
programme.

  
 \bibliographystyle{mn2e}
 \bibliography{references}

\bsp
\label{lastpage}  
\end{document}